\documentclass[preprint,preprintnumbers,amsmath,amssymb]{revtex4}

\usepackage{graphicx}
\usepackage{dcolumn}
\usepackage{bm}

 \newcommand{\lp}{\left(}
  \newcommand{\rp}{\right)}

    \newcommand{\mE}{\tilde \mu_E}
 \newcommand{\tE}{\tilde t_E}
  \newcommand{\mEs}{\mE^{(s)}}

\begin{document}


\title{Complex singularities around the QCD  critical point \\  at   finite densities}

\author{Shinji Ejiri}
 \email{ejiri@muse.sc.niigata-u.ac.jp}
\affiliation{Graduate School of Science and Technology, Niigata University,  Niigata 950-2181, Japan}
\author{Yasuhiko Shinno}%
 \email{shinno@libe.nara-k.ac.jp}
\affiliation{%
Nara National College of  Technology,Yamatokoriyama, Nara 639-1080, Japan
}%

\author{Hiroshi Yoneyama}
 \email{yoneyama@cc.saga-u.ac.jp}
\affiliation{
Department of Physics, Saga University, Saga 840-8502, Japan\\
}%

\date{\today}

\begin{abstract}
Partition function zeros   provide alternative approach to study phase structure of finite density QCD.  The structure of   the Lee-Yang edge singularities associated with the zeros  in the complex chemical potential  plane has  a strong influence   on the real axis of the chemical potential. 
   In order to investigate 
what the singularities are like  in a concrete form,  we resort to an effective theory based on a mean filed approach     in the vicinity of  the  critical  point.
 The crossover is identified as a real part of the singular point.  
  We consider the complex effective potential  and explicitly  study  the behavior of its extrema in the complex order parameter  plane   in order to see  how the Stokes lines are associated with the singularity.  Susceptibilities in the complex plane are also discussed. 
\end{abstract}
\maketitle

\section{Introduction}
The  critical  point (CP)~\cite{AY, BCCGP, SRS, HJSSV,BR} of QCD at  finite temperatures and densities is an important issue, and its existence and associated nature of  the   quark gluon plasma phase  may be clarified by the  heavy ion experiments in the near future~\cite{FH}.   This  is  one of the main targets  lattice simulations are aiming at.  The situation is,  however, not conclusive due to  the notorious sign problem.   Various approaches such as the Taylor series~\cite{AEH, AEH2, ADE, GG}  and  the imaginary chemical potential~\cite{Lo, deFP, deFP2, DL, DL2, CCDP, CCEMP} are adopted in order to circumvent the sign problem~\cite{MNNT, EKU, E0}.
Their validity is controlled by the thermodynamic singularities in the complex chemical potential plane~\cite{S, SMF}.  Such singularities are deeply associated with the partition function zeros.   
In \cite{EY}, the QCD singularities have been investigated in the complex $\mu$ plane by using $N_f=2$ QCD with staggered quarks.
 In the present paper,  we pursue this issue in terms of an effective theory, by focusing on the  singularities   in the vicinity of the CP in the complex chemical potential plane. \par
Study of the partition function zeros~\cite{YL, LY, IPZ, BMKKL,BMKKL2, ACFU, FK, E, NN,DDLMZ, LM, S,GLMS,ESY}  provides an alternative approach to the critical phenomena and its scaling behaviors.   The partition function as a function of parameter $\lambda$  such as temperature and magnetic field for finite volume takes zero at $\lambda=\lambda_k$   in the complex $\lambda$ plane.  
$$Z(\lambda)=\prod_k\left(\lambda-\lambda_k\right).$$
As  pointed out by Lee-Yang~\cite{LY}, there is   an analogy with two dimensional Coulomb gas.  In the infinite volume limit, 
the zeros  accumulate  on curves $C$ with a  ``charge  density''  $\xi(s)$, and the real part of the free energy $\Omega(\lambda)$ becomes 
\begin{eqnarray}
{\rm Re}\ \Omega(\lambda)=-T\int_C ds \  \xi(s)\log\left|\lambda-\lambda(s)\right|.
\label{eq:Stokes}
\end{eqnarray}
${\rm Re}\ \Omega$ is continuous across $C$, while the 
``electric field'' 
${\bf E}=-\nabla {\rm Re}\left(\Omega\right)$
is discontinuous in the direction normal to the curve. The amount of the discontinuity is proportional to the charge density $\xi$. 
This curve named as the Stokes line is regarded as the location of  a cut on a  Riemann sheet of the analytic function ${\rm Re}\Omega$.  It should be noted that the singularities occur at zeroes $\lambda_k$ for finite volume, while they appear  only at  branch points (not on the cut) in the infinite volume limit~\cite{S}. \par
Such branch points are  termed  the Lee-Yang edge  singularities.    The Lee-Yang edge  singularities  have    a  strong influence, as the closest singularities to the real axis,  on the behaviors of  thermodynamic quantities  for  the real values of $\lambda$.   They can also be regarded as critical points in the complex plane \cite{F}, and thermodynamic quantities become singular with the  critical exponents characterized by the Lee-Yang edge  singularities.
In order to investigate  what the   edge singularities are like in the vicinity of the CP,  we resort to an effective theory reflecting the phase structure of QCD~\cite{HI}.  This model is constructed based on  the tricritical point (TCP) in the $\mu$-T plane in the chiral limit, which has the upper critical dimension equal to 3, and thus its  mean field description  is expected to be valid  up to a logarithmic correction. It  provides some interesting physics in the vicinity of the tricritical point for vanishing quark mass $m$, and of the CP for small $m$.  We make use of  this model to investigate the nature in the complex $\mu$ plane.   
 Thermodynamic singularities in the  complex $\mu$ plane have been  studied by Stephanov in terms of the random matrix theory~\cite{S}. In the present paper,  we pay   
more   attention to  the influence of the singularities  on the real $\mu$ axis.  
  \par
When one introduces complex $\mu$, the order parameter also becomes complex and so does the effective potential.   
Its $\mu$ dependence is quite intricate in the complex case.  The  above stated model, however,   allows to analyze the complex potential. By focusing on  the real part of the potential, we explicitly trace its extrema.  In the vicinity of the singularity,  their movements 
show different behaviors depending on where  they pass.  From its behaviors, we identify where the Stokes lines are located.   We also see along the Stokes line  that  the critical exponent associated with the gap of Im~$\Omega$ around  the singular point differ from that on the real axis.  
The  chiral susceptibility and quark number susceptibility  in the complex $\mu$   plane are also  discussed. By  moving Re~$\mu$ with fixed Im~$\mu$,   the complex susceptibilities show a distinct  behavior between  Im~$\mu<$ Im~$\mu^{(s)}$ and  Im~$\mu>$ Im~$\mu^{(s)}$, where $\mu^{(s)}$  denotes the singular point in the complex $\mu$ plane.    We trace the peak of the complex susceptibilities.    It turns out that the  chiral susceptibility 
develops a peak at a location in agreement  with the real part of the singular point in the vicinity of the CP. 
As a  reminiscence of the singularity, the location of the peak of both the susceptibilities on the real 
 axis    shows the same temperature dependence  as that of  Re~$\mu^{(s)}$.   
 \par
 In the following section,  after    a brief explanation of the model,   we   see how the effective potential looks above, on  and below  temperature of the CP for various values of  real $\mu$.  We then move to the complex $\mu$ plane, and investigate the   
  singularities.  Chiral and quark number susceptibilities are discussed in connection with the singularities.    In section 3, we explicitly study the extrema of the complex effective potential and discuss the Stokes lines.  The chiral and quark number susceptibilities in the complex plane are  also discussed.    Summary is presented in section 4.
\section{Edge singularities }\label{sec:edge}
We investigate the edge singularities of QCD in the complex $\mu$ plane. For this purpose, we resort to some effective theory describing  the phase structure of QCD around the critical point.  In the present paper, we adopt  a model   proposed by Hatta-Ikeda~\cite{HI}. Although this model is based on the mean field, it is interesting in the sense that 
the inter-relationship  between the TCP and the CP indicates characteristic behaviors of the phase structure.  
\subsection{Effective potential  and CP }
In this subsection, let us briefly explain the model which we deal with in the paper.
We consider $N_f=2$ case. In the chiral  limit,  there exists a TCP  at finite temperature and density. The TCP  is connected to  a  critical point  at $\mu=0$, which is in the same universality class as 3-d  O(4) spin mode~\cite{PW}.   And a first order line goes  down  from the TCP toward lower temperature side. 
When quark mass $m$ is introduced,  the critical line is absent, and the surviving first order line  terminates at a critical point  (CP). This   CP  can be described   by fluctuations of the  sigma meson  and is expected to share the same universality with 3-d Ising model~\cite{HJSSV}.  Since the upper critical dimension of the tricritical point is equal to 3,  the TCP in    QCD phase diagram  can  be described by a mean field theory up to a  logarithmic correction. 
As far as   $m$ is small,   the universal behavior around the CP is also expected to be described in the mean field framework. Let us briefly explain the model~\cite{HI} in the following. \par
Starting with the Landau-Ginzburg potential, which  incorporates  only  the  long wavelength contribution 
\begin{eqnarray}
\Omega_{LG}=-m\sigma +\frac{a}{2}\sigma^2+\frac{b}{4}\sigma^4+\frac{c}{6}\sigma^6, 
\label{eq:TCP} 
\end{eqnarray}
one expands it around the TCP ($a=b=m=0$) assuming $a$ and $b$ as  a linear  function of $\mu$ and $T$, 
\begin{eqnarray}
a(T,\mu)=C_a \tilde{t}_3+D_a\tilde {\mu}_3, \quad
b(T,\mu)=C_b \tilde{t}_3+D_b\tilde {\mu}_3,
\label{eq:linear}
\end{eqnarray}
where $\tilde {\mu}_3=\mu-\mu_3$  ($\tilde{t}_3=T-T_3$),   and  ($ \mu_3, T_3$) denotes  temperature and chemical potential at the TCP. A coefficient $c$ in Eq.~(\ref{eq:TCP})  is assumed to be constant at the TCP. 
\begin{figure}
\vspace{-5mm}
        \centerline{\includegraphics[width=10cm, height=4
cm]{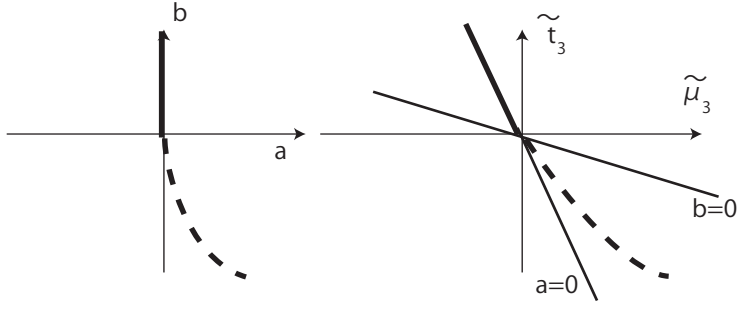}}
\caption{ Mapping of the phase diagram from $(a,b)$-plane to ($\tilde{\mu}_3, \tilde{t}_3$)-plane. Thick solid (broken) lines indicate a second order critical lines (first order phase transition lines).  }
\label{fig:map}
\end{figure}
Coefficients $C_a, D_a, C_b$  and $D_b$ in Eq.~(\ref{eq:linear}) are constrained  to some extent from the  structure around the TCP.  Positivity  of $C_a$ ($C_a>0$)  is  determined  from $a(T,\mu)>0$ (the symmetric phase) for $\tilde{t}_3>0$ at $\tilde {\mu}_3=0$, and similarly $D_a>0$ is from  $\tilde {\mu}_3>0$ at $\tilde{t}_3=0$. As illustrated in Fig.~\ref{fig:map}, 
a second order phase transition line ($a=0, b\geq 0$) in Eq.(\ref{eq:TCP}) is  mapped to a straight line $\tilde{t}_3=-(D_a/C_a)\ \tilde {\mu}_3$ ($\tilde {\mu}_3\leq 0$ and $\tilde{t}_3\geq 0$). 
 Moreover,   a first order line ($a=3b^2/(16c)$),  existing in the region  $a>0$ and $b<0$ and tangential to $a=0$   at the origin in the $a$-$b$ plane,   is expected to be mapped  into  the region    $\tilde {\mu}_3>0$ and  $\tilde {t}_3<0$ and tangential to the straight line $a=0$ at the TCP. This  is realized when  the followings hold
\begin{eqnarray}
C_bD_a-C_aD_b>0, \qquad  C_a,C_b,D_a,D_b >0, 
\label{eq:cond1}
\end{eqnarray}
where the left condition  implies  that  the linear transformation from $(a,b)$ to $(\tilde {\mu}_3, \tilde{t}_3)$ keeps the orientation. 
\par
 By switching on $m$, the condition for the CP to exist at $T=T_E$ and $\mu=\mu_E$
\begin{eqnarray}
\frac{\partial \Omega(T_{E}, \mu_{E}, \sigma_0)}{\partial \sigma}=\frac{\partial^2 \Omega(T_{E}, \mu_{E}, \sigma_0)}{\partial \sigma^2}=
\frac{\partial^3 \Omega(T_{E}, \mu_{E}, \sigma_0)}{\partial \sigma^3}=0
\end{eqnarray}
 leads  to the coefficients 
 \begin{eqnarray}
a(T_{E}, \mu_{E})=\frac{9b(T_{E}, \mu_{E})^2}{20 c}, \quad -b(T_{E}, \mu_{E})=\frac{5}{54^{1/5}}c^{3/5}m^{2/5}, 
\end{eqnarray}
and expectation value of $\sigma$
\begin{eqnarray}
 \sigma_0=\sqrt{\frac{-3 \ b(T_{E}, \mu_{E})}{10 c}}.
 \label{sig0}
\end{eqnarray}
The CP is deviated from the TCP as follows;
\begin{eqnarray}
T_{E}-T_3 &=& -\frac{5 D_a c^{3/5}}{(54)^{1/5}\left(C_bD_a-C_aD_b\right)}m^{2/5}, \\
\mu_{E}-\mu_3&=& \frac{5 C_a c^{3/5}}{(54)^{1/5}\left(C_bD_a-C_aD_b\right)}m^{2/5}.
\label{eq:CEP}
\end{eqnarray}
It is noted that with the condition Eq. (\ref{eq:cond1}), $T_{E}<T_3$ and $\mu_{E}>\mu_3$ hold.  
\par 
Expanding $\Omega(T,\mu,\sigma)$ around  $\Omega(T_{E}, \mu_{E},\sigma_0)$, we obtain 
thermodynamic potential around the CP  given by 
\begin{eqnarray}
\Omega(T,\mu,\sigma)=\Omega(T_{E}, \mu_{E}, \sigma_0)+A_1\hat {\sigma} +A_2\hat {\sigma}^2+A_3\hat {\sigma}^3+A_4\hat {\sigma}^4,
\label{eq:Omega_CEP}
\end{eqnarray}
where $\hat {\sigma}=\sigma-\sigma_0$.  This is the potential we use in the present paper. \par
The coefficients $A_i$ are given as follows as  a function of $T$ and $\mu$.
\begin{eqnarray}
A_1&=&\lp C_a\sigma_0+C_b\sigma_0^3\rp \tilde{t}_E+\lp D_a\sigma_0+D_b \sigma_0^3\rp\tilde{\mu}_E\nonumber\\
A_2&=&\frac{1}{2}\lp C_a+3C_b\sigma_0^2\rp \tilde{t}_E +\frac{1}{2}\lp D_a+3D_b\sigma_0^2\rp \tilde{\mu}_E \nonumber\\
A_3&=& \left\{C_b \tilde{t}_E +D_b\tilde{\mu}_E \right\} \sigma_0\nonumber\\
A_4&=&-\frac{b(T_{E}, \mu_{E})}{2}+\frac{1}{4}\left(C_b\tilde{t}_E  +D_b\tilde{\mu}_E\right), 
\label{eq:coeff}
\end{eqnarray}
 where $\tilde{t}_E\equiv T-T_E$ and $\tilde{\mu}_E\equiv \mu-\mu_E$. 

The stability $A_4>0$ of the potential (\ref{eq:Omega_CEP})  gives 
\begin{eqnarray}
\tilde{\mu}_{E}> \frac{2 b(T_{E}, \mu_{E})-C_b\tilde{t}_{E} }{D_b}. 
\label{eq:stability}
\end{eqnarray}
It is checked if this stability condition is fulfilled in the following  calculations. 
\par
\begin{figure}
        \centerline{\includegraphics[width=12cm, height=6
cm]{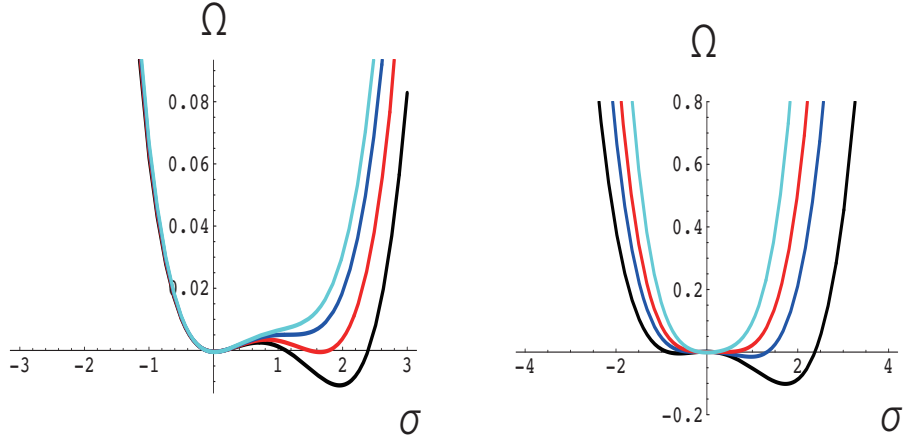}}
\caption{ Behaviors of    $\Omega$ for $\tilde{t}_E\equiv T-T_{E}< 0$  (left)  and  $\tilde{t}_E=0$ (right).  $C_a=0.1$ and $x_m=m^{1/5}=0.2$.  The other parameters are $C_b=D_a=D_b=c=1.0$. Left:  
For $\tilde{t}_E=-0.2$,     the values of $\tilde{\mu}_E\equiv  \mu-\mu_{E}$  are chosen to be 0.054, 0.0555 (red line: phase transition point), 0.058, 0.06 from bottom to top.  Right:  For $\tilde{t}_E=0$, the value of 
$\tilde{\mu}_E$ is chosen to be $ -0.1, -0.05, 0$ (red line: CP)  and 0.1.  }
\label{fig:effctive-pot-1}
\end{figure}
\begin{figure}
        \centerline{\includegraphics[width=12cm, height=6
cm]{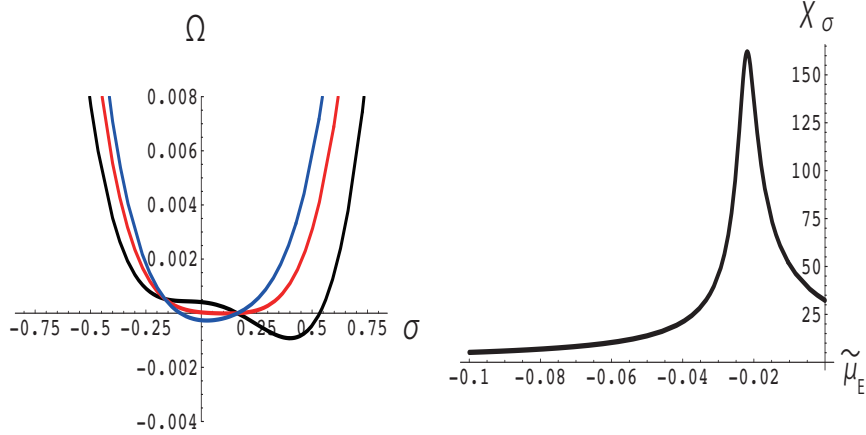}}
\caption{ Behaviors of    $\Omega$ for $\tilde{t}_E>0$  (left).   Left:  
For $\tilde{t}_E=0.2$,    the value of $\tilde{\mu}_E$ is chosen to be -0.05,  -0.0222 (red line: crossover), and -0.001.  Right: the inverse curvature  at  the global minimum of $\Omega$ as a function of $\tilde{\mu}_E$ for $\tilde{t}_E=0.2$.}
\label{fig:effctive-pot-2}
\end{figure}

Before  discussing   the properties of the model in the complex $\mu$ plane, it would be  better to see  what this model is like in the real $\mu$ case. This will  also be helpful in order to discuss the Stokes line in the following section. 
Figures \ref{fig:effctive-pot-1}  and \ref{fig:effctive-pot-2} indicate  typical  behaviors of $\Omega$ at    temperature around the CP.  
 Here we chose  the following numerical values for simplicity,  
 \begin{eqnarray}
x_m=m^{1/5}=0.2, \quad  C_a=0.1, \quad C_b=D_a=D_b=c=1.0. 
\label{eq:parameters}
\end{eqnarray}
and the same  values for  these parameters  are used for the calculations throughout the paper. 
The left panel of Fig.~\ref{fig:effctive-pot-1} indicates the behavior of a first order phase transition for a negative  value of $\tilde{t}_E (=-0.2)$.  The values of $\tilde{\mu}_E\equiv  \mu-\mu_{E}$  are chosen to be  0.054, 0.0555, 0.058, 0.06 from bottom to top. Note that at $\tilde{t}_E=-0.2 $, a first order phase transition occurs at  $\tilde{\mu}_E=0.0555$ (red line).  In the right panel, the behaviors at  $\tilde{t}_E=0$ are shown ($\tilde{\mu}_E$ is chosen to be $  -0.1, -0.05, 0$ and 0.1 from bottom to top). The critical point corresponds to $\tilde{\mu}_E=0$ (red line).\\
 In   the left panel of Fig.~\ref{fig:effctive-pot-2}, behaviors of    $\Omega$ for positive value of $\tilde{t}_E (=0.2)$  are shown,  where $\tilde{\mu}_E$ is chosen to be -0.05, -0.0222, and -0.001.
 At $\tilde{t}_E=0.2$, $\Omega$ develops an  approximately flat minimum  at $\tilde{\mu}_E=-0.0222$ (red line).  The right panel indicates   the inverse curvature  of $\Omega$ at  the global minimum  as a function of $\tilde{\mu}_E$,    which develops   a peak at  $\tilde{\mu}_E=-0.0222$.  

 \subsection{Edge singularities}
 Let us now move to the complex $\mu$ plane.  Using the potential in Eq.~(\ref{eq:Omega_CEP}), 
 the instability of the extrema occurs at such $\sigma$  that
 \begin{eqnarray}
\frac{\partial \Omega}{\partial \sigma }=0, \quad \frac{\partial^2 \Omega}{\partial \sigma^2 }=0
\label{eq:singularity}
\end{eqnarray}
are simultaneously satisfied.   Namely, 
\begin{eqnarray*}
A_1 +2 A_2\sigma+3 A_3\sigma^2+4 A_4\sigma^3=0, \quad 2 A_2+6 A_3\sigma+12 A_4\sigma^2=0.
\end{eqnarray*}
This occurs when the former cubic equation  has vanishing discriminant.
\begin{eqnarray}
-4 a_1^3 a_3 + a_1^2 a_2^2 - 4 a_0 a_2^3 + 18 a_0 a_1 a_2 a_3 - 27 a_0^2 a_3^2 =0,
\label{eq:disc}
\end{eqnarray}
where
\begin{eqnarray}
a_0 = 4A_4, \quad a_1 = 3A_3, \quad  a_2 = 2 A_2, \quad a_3 = A_1.
\end{eqnarray}
Using the coefficients $A_i$ in Eq.~(\ref{eq:coeff}), the discriminant Eq.~(\ref{eq:disc}) is solved as a function of $\tilde\mu_E$ and $\tilde t_E$.
 \par
\begin{figure}
        \centerline{\includegraphics[width=6cm, height=6
cm]{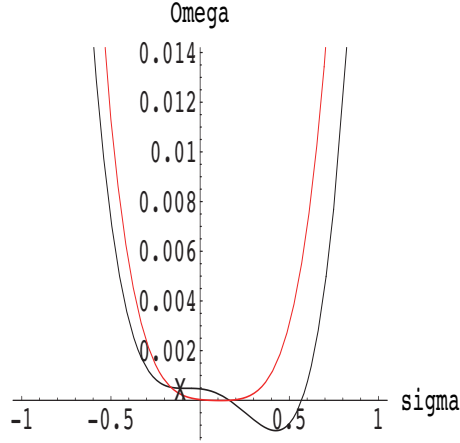}}
\caption{Potential $\Omega$, Eq.(\ref{eq:Omega_CEP}),  for $\tilde{t}_E=0.2$.  Two values ${\tilde \mu}_E$ are chosen.  Black line: ${\tilde \mu}_E=-0.0538$, where  $\Omega^{'}=\Omega^{"}=0$ is fulfilled at $\sigma=-0.07602$ ($\times$).   Red line: ${\tilde \mu}_E=-0.0222$, which is equivalent to the real part of (iii) in Eq.(\ref{eq: soltodisc}).
 }
\label{fig:pot-ttildE02}
\end{figure}

 
 For  ${\tilde t}_E=0.2$ (Fig.~\ref{fig:effctive-pot-2}), for example,  the discriminant Eq. (\ref{eq:disc}) yields four roots such as 
\begin{eqnarray}
{\tilde \mu}_E=-0.3857\  ({\rm i}), \quad -0.0538 \  ({\rm ii}), \quad -0.0222\pm 0.00254 \ i \  ({\rm iii}). 
\label{eq: soltodisc}
\end{eqnarray}
  The stability condition of the potential Eq.(\ref{eq:stability}) gives  ${\tilde \mu}_E>-0.3801$ for ${\tilde t}_E=0.2$, and thus the solution (i) is excluded. In  the case (ii),  
  an inflection point is located   
 at $\sigma=-0.07602$.   This point  has  nothing to do with the phase transition\footnote{Actually, it corresponds to one of the extrema as will be discussed in subsection \ref{subsec:AC}, depicted as (C) in the right panel of Fig.~\ref{fig:Stokes-1}.  Since this point  is misleading from the viewpoint of the singularity,  it   is stressed as   shown  in Fig.~\ref{fig:pot-ttildE02} (cross symbol on the black line). 
  This extremum does not take part in the phase transition.}. The remaining solutions (iii),  with their  real part satisfying the stability condition,    can be  identified as   a pair of  complex singularities at ${\tilde t}_E=0.2$  associated with the  CP at ${\tilde t}_E=0$.   Singularities of type (iii) are denoted by  ${\tilde \mu}^{(s)}_E$, and we focus on these hereafter.   
 \par
\begin{figure}
        \centerline{\includegraphics[width=14cm, height=6
cm]{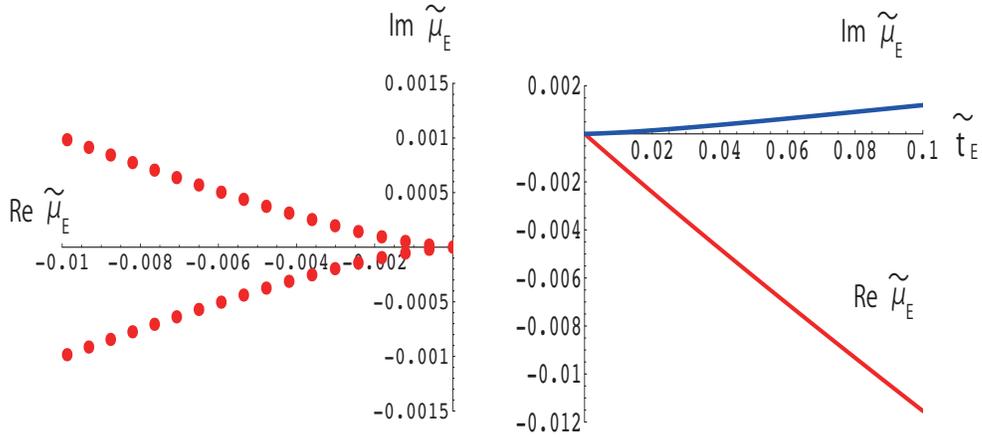}}
\caption{Left: Singularities in the complex $\tilde{\mu}_E=\mu-\mu_{E}$ plane. Temperatures are chosen to be  $\tilde{t}_E= T-T_{E}=0, 0.005, 0.01, 0.15, 0.02 \dots  0.09$.  Right:    The $\tilde{t}_E$-dependence of Re $\tilde{\mu}_E^{(s)}$ (red) and  Im $\tilde{\mu}_E^{(s)}$ (blue).  Re $\tilde{\mu}_E^{(s)}$ is linear in $\tilde{t}_E$, while  Im $\tilde{\mu}_E^{(s)}$ behaves  as $\tilde{t}_E^{3/2}$ \cite{ESY}. 
 } 
\label{fig:edge-sing}
\end{figure}

The left panel of Fig.~\ref{fig:edge-sing} indicates the locations of  ${\tilde \mu}^{(s)}_E$ in the complex $\tilde{\mu}_E$  plane  for various values of temperature ($\tilde{t}_E\geq 0$). Temperatures are chosen to be  $\tilde{t}_E= T-T_{E}=0, 0.005, 0.01, 0.15, 0.02 \dots  0.09$.    At $\tilde{t}_E=0$, the singularity of the CP appears on the real $\tilde{\mu}_E$ axis (the origin in the figure).
For  $\tilde{t}_E> 0$,    edge singularities appearing in pairs deviate from the real $\tilde{\mu}_E$-axis as $\tilde{t}_E$ increases. The $\tilde{t}_E$-dependence of Re $\tilde{\mu}_E^{(s)}$ and Im  $\tilde{\mu}_E^{(s)}$ is shown in the right panel.   The real part depends  linearly on $\tilde{t}_E$, while the imaginary one behaves as $\tilde{t}_E^{\beta\delta}$ with $\beta\delta=3/2$ as expected~\cite{S}.
\par
 In comparison to  the result  in Ref.\cite{S},   we translate the behavior of the singularities in the $\tilde{\mu}_E$ plane to  that  in the $\mu^2$ plane. For this  we need the location of the tricritical point,   which is however unknown in this frame work.   
By putting some appropriate numbers for the tricritical point $\mu_3$, we could  plot the  singularities in the complex $\mu^2$ plane.
  ${\rm Re}\ \mu={\rm Re}\ \tilde{\mu}_E+\mu_0$ and ${\rm Im}\  \mu={\rm Im}\  \tilde{\mu}_E$. Here,  $\mu_0$ is given in terms of the coefficients $C_i$'s and $\mu_3$;  
  \begin{eqnarray}
\mu_0=\frac{5C_ac^{3/5}}{(54)^{1/5}(C_bD_a - C_aD_b)}m^{2/5}+\mu_3.
\end{eqnarray}
 The  singularities  deviate from the real $\mu^2$ axis as $\tilde{t}_E$ increases from 0  in the same way as  those  in Fig.~\ref{fig:edge-sing}, since a mapping from  $\tilde{\mu}_E$ to $\mu^2$ is a conformal one.  
  \subsection{Crossover}
  \label{cross}
\begin{figure}
        \centerline{\includegraphics[width=12cm, height=6
cm]{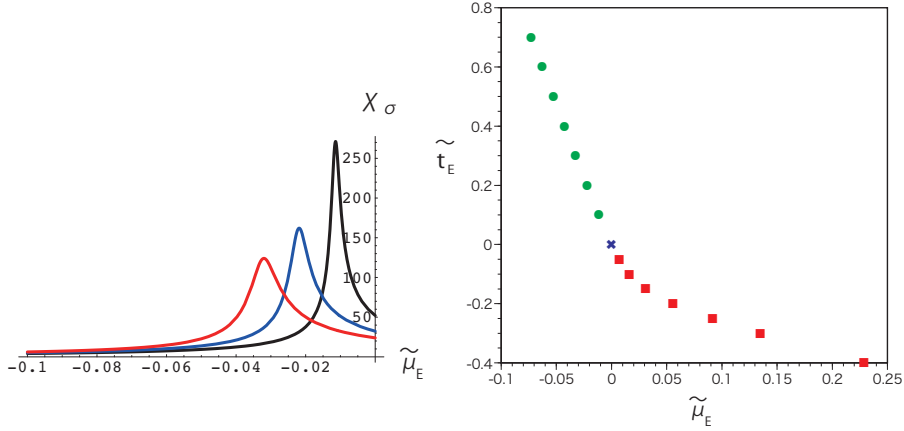}}
\caption{Left: $\chi_\sigma$  as a  function of ${\tilde \mu}_E$. $\tilde{t}_E=$0.1(black), 0.2 (blue) and 0.3 (red).  Right: Phase diagram around the CP.  The horizontal (vertical ) axis is ${\tilde \mu}_E$ (${\tilde t}_E$), and the origin ($\times$) is the location of the CP. $x_m=0.2$.   Red symbols  indicate    first order phase transition points, and green  ones do the locations of the crossover, which are in agreement with   the real parts of the singularities, Re ${\tilde \mu}^{(s)}_E$. 
 }
\label{fig: susp-xm02ttild}
\end{figure}

  Since the singular points  move away from the real $\mu$ axis for positive values of ${\tilde t}_E$,  substantial quantities  like the  chiral susceptibility show no singular behavior on the real axis, but the reminiscence of the singularity appears as  crossover.   
  At temperatures   close to the CP (${\tilde t}_E \gtrsim 0$),  it would then be  natural to identify   ${\rm Re}\ {\tilde \mu}^{(s)}_E$ as the location of the crossover  on the real   $\mu$ axis~\cite{S}.  In the  example, Eq. (\ref{eq: soltodisc}), at  ${\tilde t}_E=0.2$,   
  $\Omega$ for ${\tilde \mu}_E= {\rm Re}\ {\tilde \mu}^{(s)}_E=-0.0222$   looks as shown in  Fig.~\ref{fig:pot-ttildE02} (red line), and   the chiral susceptibility  $\chi_\sigma$ becomes maximal at  the value in agreement with   $ {\rm Re}\ {\tilde \mu}^{(s)}_E$  as was shown in the right panel of Fig.~\ref{fig:effctive-pot-2}, where  $\chi_\sigma$ is  given by inverse curvature of the effective potential $  \Omega(\sigma)$ at the global minimum $\bar \sigma$
  \begin{eqnarray}
\chi_\sigma=-\frac{\partial^2  \Omega(\bar\sigma(m), m)}{\partial m^2}=\left. \frac{1}{\frac{\partial^2  \Omega(\sigma)}{\partial \sigma^2}}\right|_{\bar \sigma}. 
\end{eqnarray}
   The  ${\tilde t}_E$-dependence of $\chi_\sigma$ as a function of ${\tilde \mu}_E$ is plotted in the left panel of Fig.~\ref{fig: susp-xm02ttild}. The peak of the curve shifts away from ${\tilde \mu}_E=0$ as ${\tilde t}_E$ increases, and its  location   are   in agreement with the values of ${\rm Re}{\tilde \mu}^{(s)}_E$ for each ${\tilde t}_E$.  
 \par 
The right panel of Fig.~\ref{fig: susp-xm02ttild} indicates the locations of the thus identified crossover for various values of temperature ($\tilde{t}_E>0$), 
together with the locations of the first order phase transition for $\tilde{t}_E<0$.  
 Red symbols  indicate    first order phase transition points,  green ones do the locations of the crossover, and the CP is located at the origin.
\subsection{Quark  number susceptibility}
\begin{figure}
        \centerline{\includegraphics[width=7cm, height=5
cm]{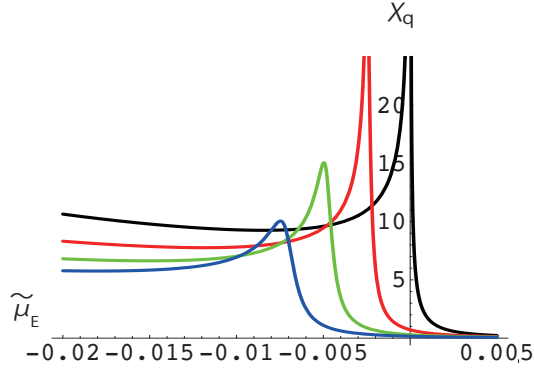}}
\caption{Quark number susceptibility  as a function of $\tilde{\mu}_E$ for four values of temperature   ($\tilde{t}_E=$0 (black), 0.02 (red), 0.04 (green)  and 0.06 (blue)). 
}
\label{fig:quark-sus}
\end{figure}

At the CP, fluctuations of the quark number as well as the heat capacity become singular in the same exponent  as that of the chiral susceptibility~\cite{HI, FO}.  
  For $\tilde t_E>0$, the singularity in the complex plane makes an effect on the locations of the crossover  in a slightly different way.  
The quark  number susceptibility is defined by 
 \begin{eqnarray}
\chi_q=-\frac{\partial^2 \Omega(\bar\sigma, \mu)}{\partial \mu^2}, 
\end{eqnarray}
 where  $\bar\sigma$ is the value of $\sigma$ at the global minimum of the potential  in (\ref{eq:Omega_CEP}).   
Figure \ref{fig:quark-sus} shows $\chi_q$ 
as a function of  $\tilde{\mu}_E$ ($\tilde{t}_E=$0 (black), 0.02 (red), 0.04 (green) and 0.06 (blue) ).
 At the CP ($\tilde{t}_E=0$),  $\chi_q$ diverges at $\tilde{\mu}_E=0$, and away from the CP ($\tilde{t}_E>0$),  $\chi_q$ develops a finite amount of peak, whose  hight becomes smaller as $\tilde{t}_E$ increases.   
Locations of the peak of  $\chi_q$ and $\chi_\sigma$  are approximately the same, but  deviate from each other as temperature increases from the CP temperature, reflecting the difference of explicit dependence of   $\Omega^{'}=\partial \Omega/\partial \sigma$ on $m$ and $\mu$, respectively; 
\begin{eqnarray}
\chi_\sigma= \frac{\left(\frac{\partial \Omega^{'}}{\partial m}\right)^2}{\frac{\partial^2  \Omega(\bar \sigma)}{\partial \sigma^2}} =\frac{1}{\frac{\partial^2  \Omega(\bar\sigma)}{\partial \sigma^2}} , \quad
 \chi_q= \frac{\left(\frac{\partial \Omega^{'}}{\partial \mu}\right)^2}{\frac{\partial^2  \Omega(\bar\sigma)}{\partial \sigma^2}}. 
\end{eqnarray}
This will be discussed again in connection with the complex susceptibilities in \ref{complexsus}. %
\section{Analytic continuations}
\label{sec:Stokes}
In this section, we consider the analytic continuation of the extrema of the effective potential. 
 By tracing them in the complex  $\mu$ plane,   the  Stokes lines are identified, which  reflects the analytic structure around  the branch points.   It will  also be seen that  the crossover phenomenon of  $\chi_\sigma$ on the real $\mu$ axis reflects  the characteristics of the singular behavior of   $\chi_\sigma$  in the complex plane. It is understood that although  we focus on the behaviors  in  the complex upper-half $\mE$ plane, 
  those also occur   in the lower-half  plane  in the  complex conjugate manner. 
\begin{figure}
        \centerline{\includegraphics[width=13cm, height=5
cm]{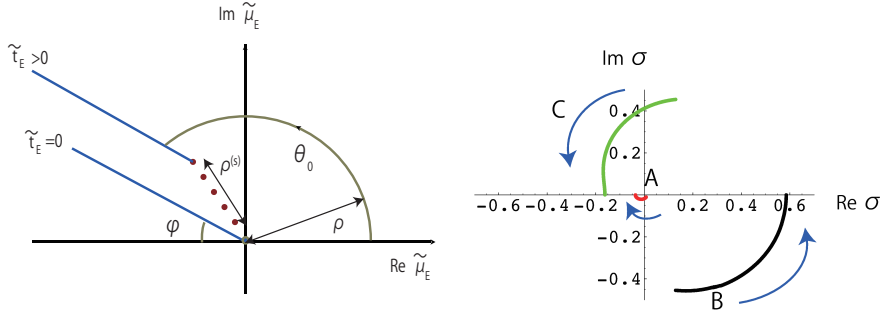}}
\caption{ Left: Stokes line (blue) in the complex   $\tilde{\mu}_E$ plane.  The singular point $\tilde{\mu}_E^{(s)}$ for  $\tilde{t}_E=0$ is located at the origin, while for  $\tilde{t}_E>0$ one of complex pair of $\tilde{\mu}_E^{(s)}$ is shifted to  the negative ${\rm Re }~\tilde{\mu}_E$ direction in the upper half plane.   Angle $\varphi$ is defined from the negative side of ${\rm Re }~\tilde{\mu}_E$ axis.  See text for detail. Right: Behaviors of   three extrema  of ${\rm Re}\ \Omega$ in the complex $\sigma$ plane  as   $\tilde{\mu}_E=\rho e^{i \theta}$ varies from  $\theta=0$ to $\theta=\pi$ ($\rho=0.05$) \cite{ESY}.}
  \label{fig:Stokes-1}
\end{figure}
\subsection{Analytic continuation of extrema of the effective potential}\label{subsec:AC}
The Stokes line is understood as a curve to which  the Lee-Yang zeros accumulate as shown in  Eq. (\ref{eq:Stokes}).  
When the phase transition occurs,  zeros accumulate onto the critical point by which two phases  are separated on the real axis.  As the parameter is   analytically continued from the one phase to the other, the corresponding global minimum of the potential  is also analytically continued.  In the present system, 
for  $\tilde{t}_E=0$, the critical point   exists at    $\tilde{\mu}_E=0$,  
while for  $\tilde{t}_E >0$,  
 the singular point  is shifted into the complex plane. 

 \par
\begin{figure}
        \centerline{\includegraphics[width=12cm, height=4
cm]{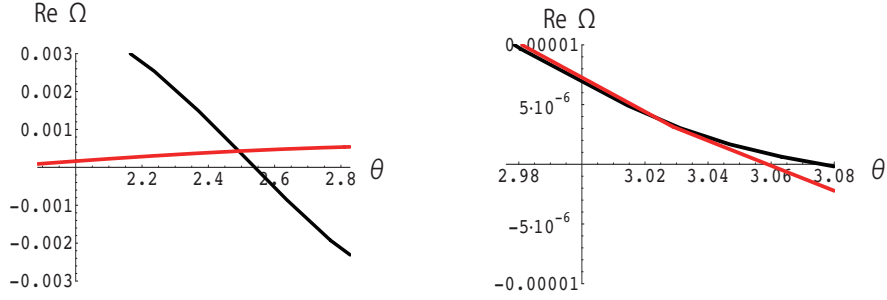}}
\caption{ Behaviors of    ${\rm Re}\ \Omega$  for the two extrema A (red) and B (black) as   $\tilde{\mu}_E=\rho e^{i \theta}$ varies  for fixed $\rho$.  $\tilde{t}_E=0.1$ and  $\rho^{(s)}=0.0116$.   Left: $\rho=0.05$.  The two values of ${\rm Re}\ \Omega$ agree at $\theta_0=2.492$.  Right: $\rho=\rho^{(s)}$. The two values of ${\rm Re}\ \Omega$ agree at  $\theta_0=3.038$.  Only the regions around $\theta_0$  are shown.}
\label{fig:real-Omega}
\end{figure}

Let  us consider  $\tilde{t}_E>0$,  and  move  $\tilde{\mu}_E$,  as shown in  the left panel of Fig.~\ref{fig:Stokes-1},  from a point on the real axis ($\tilde{\mu}_E>0$) 
 to the other side  at $\tilde{\mu}_E<0$  by changing $\theta$ from 0 to $\pi$ in $\tilde{\mu}_E=\rho e^{i \theta}$ for a fixed value of $\rho$. 
The singular point is located at $\tilde{\mu}_E^{(s)}=\rho^{(s)} e^{i \theta^{(s)}}$.
Firstly, we    choose  $\rho>\rho^{(s)}$.  
In the course of the variation of  $\theta$, a minimum of ${\rm Re}\ \Omega(\sigma)$  analytically continued   from  $\tilde{\mu}_E>0$ jumps    discontinuously,   at some value of $\theta$ ($\equiv\theta_0$),  to  the  one  continued   from  the other side of $\tilde{\mu}_E<0$ in the complex $\sigma$ plane.  
The location where  such transition occurs  corresponds to crossing the Stokes line in the complex $\tilde{\mu}_E$ plane.  From the argument in Introduction, the two values of  ${\rm Re}\ \Omega(\sigma)$   become  equal at this point. 
That is, the Stokes lines are given by
\begin{eqnarray}
{\rm Re}\  \Omega(\sigma_+)&=&{\rm Re}\  \Omega(\sigma_-),
 \label{eq2}
\end{eqnarray}
where $\sigma_+ (\sigma_-)$ is the location of the extremum analytically continued from $\tilde{\mu}_E>0$ ($\tilde{\mu}_E<0$ ).
 In order to explicitly see how this occurs, we take  $\tilde{t}_E=0.1$. At this temperature,   the singular points  are located at $\tilde{\mu}_E^{(s)}=-0.0115 \pm i \ 0.0012$ and thus $\rho^{(s)}=0.0116$. The right panel of Fig.~\ref{fig:Stokes-1} shows behaviors of   three extrema  of ${\rm Re}\ \Omega$ in the complex $\sigma$ plane as  $\theta$ varies from  $0$ to $\pi$ with fixed value of $\rho$ $(=0.05)$.   For $\theta=0$, ${\rm Re}\ \Omega$  develops three   extrema  at $\sigma=0.0108$ (A) 
and  $\sigma=0.12910-i \ 0.4544$  (B) and $0.12910+i\  0.4544$ (C), each of which is  located at the initial point of the arrows denoted by A, B and C, respectively.   
The arrows in the figure indicate the direction of the movement  of the three as $\theta$ varies from 0 to $\pi$. 
The extremum  A, which is a global minimum for $\theta=0$,  ends up with  negative small  value of  $\sigma=-0.0366$ at $\theta=\pi$, while 
 B does  with the real axis at  $\sigma=0.5848$,  which is identified as the  global minimum of $\Omega$ for $\theta=\pi$ from its shape (see Fig.~\ref{fig:effctive-pot-2}).  
  So the two extrema  A and B are associated with the global minimum of  $\Omega$  for real $\mu$, the phase for  $\tilde{\mu}_E>0$   and the  one for $\tilde{\mu}_E<0$ , respectively, while the  extremum C  is not associated  with the phase transition. 
In the left panel in Fig.~\ref{fig:real-Omega}, the behavior of ${\rm Re}\ \Omega$     for the two extrema A and B is shown as    a function of  $\theta$.  
The two values agree at $\theta_0=2.492$, where their slopes show a discontinuity  in accordance with the analogy of  two dimensional  Coulomb gas. 
\par
\begin{figure}
        \centerline{\includegraphics[width=15cm, height=5
cm]{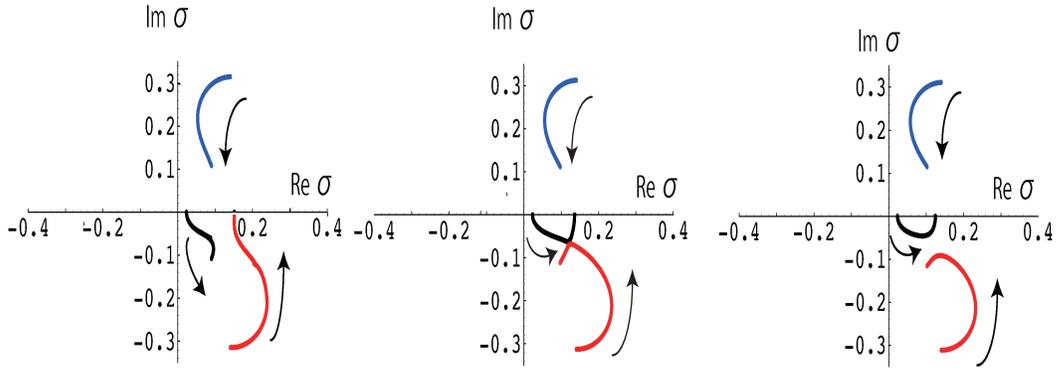}}
\caption{ Behaviors of   the three extrema  around the singular point of ${\rm Re}\ \Omega$ in the complex $\sigma$ plane  as   $\tilde{\mu}_E=\rho e^{i \theta}$ varies from  $\theta=0$ to $\pi$.  Left: $\rho=0.012$, middle: $\rho=\rho^{(s)}=0.0116$, right: $\rho=0.0113$.  Temperature $\tilde{t}_E$ is set to 0.1.  The minimum corresponding to the singularity at $\tilde{\mu}_E=-0.0115 +i \ 0.0012$,   fulfilling the condition Eq. (\ref{eq:singularity}),   is attained at $\sigma=0.1219- i \ 0.0653\equiv \sigma^{(s)}$, where two trajectories meet together in the middle panel.  }
\label{fig:extrema-traj-ttil0.1-rho}
\end{figure}
\begin{figure}
        \centerline{\includegraphics[width=16cm, height=6
cm]{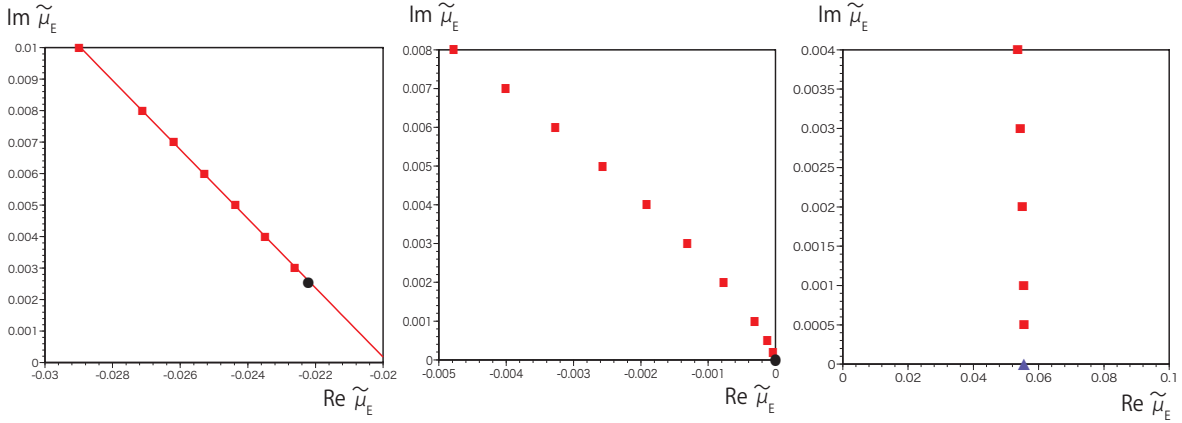}}
\caption{ Locations of the Stokes line (red) in complex $\tilde{\mu}_E$ plane, which are calculated   by ${\rm Re}\ \Omega$  for the two extrema.    Left:  $\tilde{t}_E$ is chosen to be 0.2.  Black symbol denotes the location of the singular point $\tilde{\mu}_E^{(s)}$=-0.02221+$i$ 0.00254.  It is clearly seen that thus identified points sit on a straight line with the slope $-1.09\pm 0.02$. Middle: $\tilde{t}_E=0$.  Singularity corresponding the CP is situated at the origin.  In this case the Stokes line emanates from the origin with $\varphi=\pi/2$.  Right: $\tilde{t}_E=-0.2$. A first order phase transition is located at $\tilde{\mu}_E=0.05553$ (blue triangle).  From this point, a Stokes line goes out  upright ($\varphi=\pi/2$). 
}
\label{fig:Stokes-ttil00-02}
\end{figure}
It is then  interesting to see how the trajectories  of the extrema of  ${\rm Re}\ \Omega$ change in the vicinity of  the singular point.   
For $\tilde{t}_E=0.1$, the minimum of  ${\rm Re}\ \Omega$ for $\tilde{\mu}_E$ at the  singularity, $\tilde{\mu}_E^{(s)}=-0.0115 +i \ 0.0012$,  i.e.,  fulfilling the condition Eq. (\ref{eq:singularity}),   is attained at $\sigma=0.1219- i \ 0.0653\equiv \sigma^{(s)}$. 
 Figure \ref{fig:extrema-traj-ttil0.1-rho} shows the  behaviors of   the three extrema  of ${\rm Re}\ \Omega$ in the complex $\sigma$ plane for  three different values of $\rho$; $\rho=0.012 \ (>\rho^{(s)})$, $0.0116 \ (=\rho^{(s)} )$ and $0.0113 \ (<\rho^{(s)})$.  For   $\rho=\rho^{(s)}$,   the two minima  A and B 
 approach   each other and meet together at $\sigma^{(s)}$ as shown in the middle panel of Fig.\ref{fig:extrema-traj-ttil0.1-rho}.   In this case,  ${\rm Re}\ \Omega$ of the two trajectories behaves as shown in the right panel of Fig.~\ref{fig:real-Omega}. 
  A smooth encounter of the two trajectories occurs    at $\theta_0=3.0384 \ (=0.9672\pi)$, whose value is indeed  in agreement with the location of the singularity $\tilde{\mu}_E^{(s)}=-0.0115 +i \ 0.0012$. \par
     As $\rho$ varies passing $\rho^{(s)}$     near $\tilde{\mu}_E^{(s)}$,   the two trajectories make  a rearrangement.  In the leftmost  panel of Fig.\ref{fig:extrema-traj-ttil0.1-rho} ($\rho>\rho^{(s)}$), three trajectories behave like those in the right panel of Fig.~\ref{fig:Stokes-1},  i.e.,  A (B) is the global minimum for $\tilde{\mu}_E>0$ ( $\tilde{\mu}_E<0$).  In contrast with this, for $\rho=0.012 \ (<\rho^{(s)})$,  only a single extremum A is associated with analytic continuation from $\tilde{\mu}_E>0$ to $\tilde{\mu}_E<0$, as shown    in the rightmost figure in Fig.\ref{fig:extrema-traj-ttil0.1-rho}.  That is, no encounter with the Stokes line is found in this case.  \par
     As another  case of  $\rho >\rho^{(s)}$, 
      let us comment a behavior  when  $\rho$ is chosen to be a specific value $\rho=0.04322$.
   In this case,   two extrema B and C (instead of A and B) meet together at $\sigma=-0.0795229$ for  $\theta=\pi$ ($ \tilde\mu_E=-0.04322$), where    the condition  $\Omega^{'}=\Omega^{"}=0$ is  fulfilled for $\tilde t_E$=0.1.  
    This corresponds to the extremum ($\times$) irrelevant to the  phase transition as discussed in Fig.~\ref{fig:pot-ttildE02} and the footnote.
\subsection{Stokes lines}
In the case of $\tilde{t}_E>0$, therefore, the Stokes line runs for $\rho>\rho^{(s)}$.  
   The left panel of Fig.~\ref{fig:Stokes-ttil00-02} indicates how the Stokes line emanates from the singular point $\tilde{\mu}_E^{(s)}$ in the complex   $\tilde{\mu}_E$ plane. Red filled symbols indicate the locations which are  calculated in the way  described above, i.e., by choosing several value of $\rho$, tracing the trajectories of the extrema and  checking the behaviors of Re $\Omega$.    For  $\tilde{t}_E=0.2$,  the singular point is located at $\tilde{\mu}_E^{(s)}$=-0.02221+$i$ 0.00254 (black symbol in the Figure).  It is seen  that the  Stokes line is alined on a straight line, which   is  tilted with  angle   $\varphi=\pi/4$ from  an axis parallel to the  negative axis of ${\rm Re} \tilde{\mu}_E$.  This is in agreement with analytic consideration shown in  Appendix \ref{ap1}~\footnote{In Appendix \ref{ap1}, the critical line for $\tilde t_3 >0$ in the $m=0$ case is discussed.  When $m$ is small, it is expected that  the Stokes line is tilted with the same angle $\varphi $. }.   
 \par

 As $\tilde{t}_E$ decreases to 0, $\tilde{\mu}_E^{(s)}$ approaches the origin, and  the angle    $\varphi$   shows a increasing tendency.      
  The middle panel of Fig.~\ref{fig:Stokes-ttil00-02} shows the Stokes line emanating from the origin 
  for $\tilde{t}_E=0$.  This behavior  is in agreement with $\varphi=\pi/2$ 
   as shown in Appendix \ref{ap3}.  
  For $\tilde{t}_E<0$, a first order phase transition is located on  the positive $\tilde{\mu}_E$ axis.   In the right  panel of Fig.~\ref{fig:Stokes-ttil00-02}, we show the case  for   $\tilde{t}_E=-0.2$, where
 a first order phase transition is located at $\tilde{\mu}_E=0.05553$ (blue triangle).  From this point, a Stokes line goes out  upright with $\varphi=\pi/2$  (see Appendix \ref{ap2}~\footnote{ The Stokes line is tilted with the same angle $\varphi $  as far as $m$ is small. }).  In recent Monte Carlo study~\cite{NMNNS} of low temperature and high density QCD,   the distribution of  the Lee-Yang zeros have been calculated, and it   looks similar to the  behavior in the right panel of Fig.~\ref{fig:Stokes-ttil00-02},  suggesting  a possible first order phase transition. \par
\begin{figure}
        \centerline{\includegraphics[width=12cm, height=6
cm]{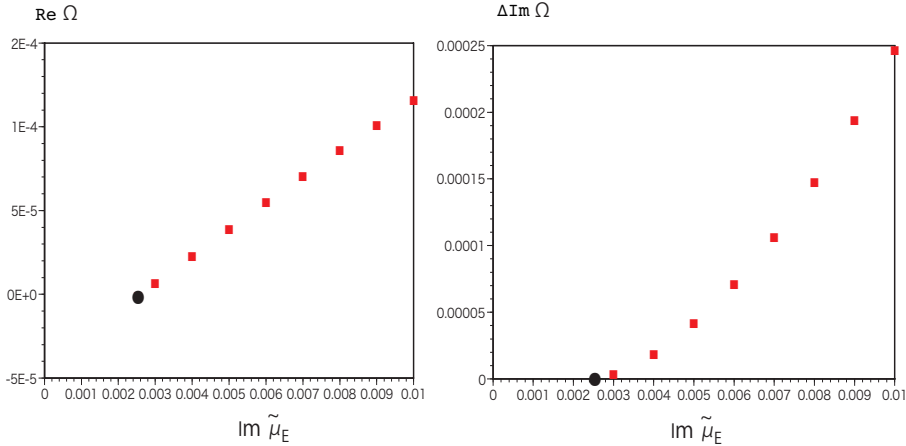}}
\caption{Re~$\Omega$ (left) and $\Delta    {\rm Im}~\Omega$ (right) as a function of Im~$\mE$ along the Stokes line for $\tE =0.2$. Black filled symbol indicates the singular point.  Re~$\Omega$ increases linearly with Im~$\mE$ (left), and  $\Delta    {\rm Im}~\Omega$ vanishes at the singular point (right).}
\label{ReOmega-ImOmega=Stokes}
\end{figure}

 
To close this subsection, we briefly discuss  Re~$\Omega$ and Im~$\Omega$ along  the Stokes line.   On the Stokes line, two values of Re~$\Omega$ continued  from  real $\mE$ axis  agree (Eq.~(\ref{eq2})).
 In the left panel of  Fig.~\ref{ReOmega-ImOmega=Stokes},  such  Re~$\Omega$ is plotted as a function of Im~$\mE$ for $\tE =0.2$, which increases linearly.   
  In contrast to this, Im~$\Omega$ shows a gap on the Stokes line (at the singular point, the gap vanishes).  
The behavior of the discontinuity $\Delta    {\rm Im}~\Omega\equiv  {\rm Im}~\Omega (\sigma_+)- {\rm Im}~\Omega (\sigma_-)$ is shown  in the right panel of  Fig.~\ref{ReOmega-ImOmega=Stokes}. The gap   grows  as 
\begin{eqnarray}
 \Delta    {\rm Im}~\Omega    \sim   ({\rm Im}~\mE-{\rm Im}~\mEs)^{3/2},
\end{eqnarray}
where ${\rm Im}~\mE$ denotes the imaginary part of the points on the Stokes line (see the left panel in Fig.~\ref{fig:Stokes-ttil00-02}).
The exponent $3/2$ comes from  the behavior of $\sigma_{+}$  near the singularity 
Re~$\sigma_{+} \sim   ({\rm Im}~\mE-{\rm Im}~\mEs)^{1/2}$, which reflects  the   fact that the critical exponents for  the edge singularity  are,   in general,      different from the usual ones  on the real axis \cite{F, IPZ}.   In the case of  $\tE=0$, we have Re~$\sigma_{+} \sim   ({\rm Im}~\mE)^{1/3}$ and $ \Delta    {\rm Im}~\Omega\sim   ({\rm Im}~\mE)^{4/3} $  (see Appendix \ref{ap-sus}). This corresponds in the     Ising case to  the magnetization behaving  like $(h-h^{(s)})^{1/2}$ near the singular point $h^{(s)}$ on the imaginary magnetic field $h$ axis for $T>T_c$  and     $\sim h^{1/\delta}$ for  $T=T_c$ ($\delta=3$ for the mean field case).  
It is also noted that at a Lee-Yang zero for finite volume, the discontinuity is like $\Delta    {\rm Im}~\Omega =(2k+1)\pi \ (k:\ {\rm integer})$ per  volume.
 \subsection{Susceptibility in the complex plane}
 \label{complexsus}
  \begin{figure}
        \centerline{\includegraphics[width=13cm, height=5
cm]{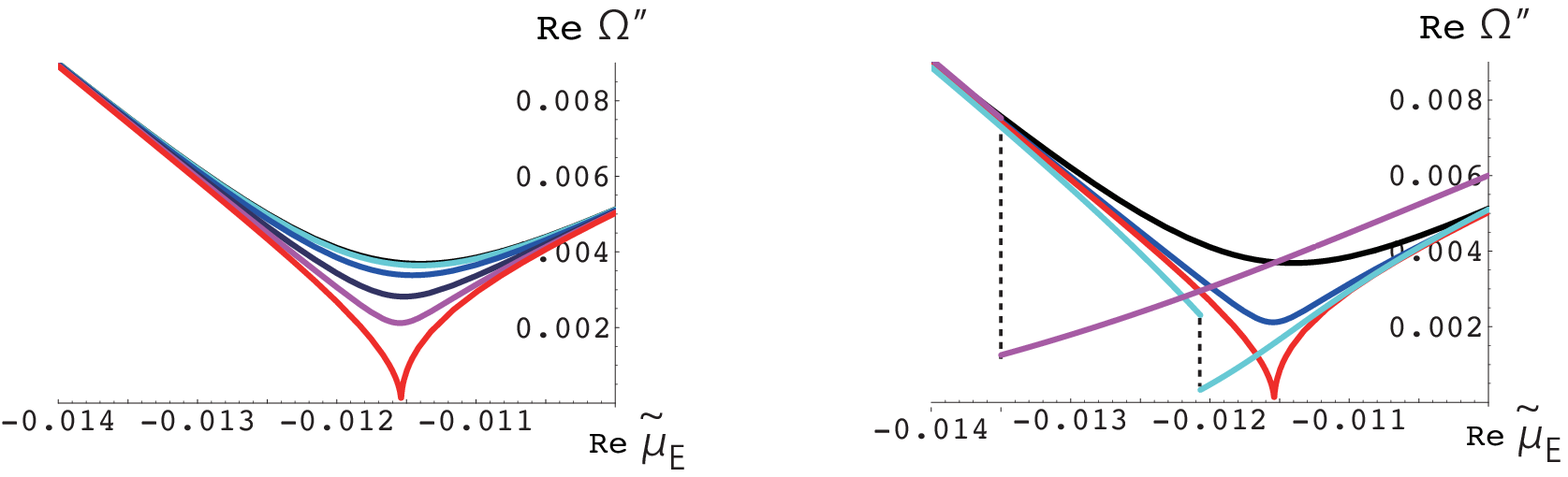}}
\caption{Behaviors of Re $\Omega"$ as a   function of Re~$\mE$ for fixed values of Im $\mE$ for $\tE=0.1$, where $\mEs=-0.011538+i \ 0.0011946$.   Left: Im~$\mE=0, 0.0002, 0.0005, 0.001$ and  $0.0011946(={\rm Im}\ \mEs$). Right: Im~$\mE=0, 0.001, 0.0011946 (={\rm Im}~\mEs), 0.002$ and  $0.004$.  For ${\rm Im}~\mE > {\rm Im}~\mEs$, Re $\Omega"$ shows a discontinuity (broken lines) when Re~$\mE$ crosses the Stokes line. }
\label{der2Omega-t01-fiximagall}
\end{figure}
The crossover behaviors of the susceptibilities on the real $\mE$ axis reflects the structure of the singularity in the complex plane.  Let us then study the behaviors of the susceptibilities in the complex plane.  We fix Im $\mE$ and move Re $\mE$ by separating the region into two, one is $0\leq$ Im~$\mE \leq$ Im~$\mEs$ and the other Im~$\mEs$ $\leq$ Im~$\mE$.  In this subsection, we plot  Re~$\Omega"=({\rm Re}~\chi_\sigma)^{-1}$ and $({\rm Re}~\chi_q)^{-1}$ so that they become vanishing at the singular point $\mEs$,   rather than diverging $\chi_\sigma$ and $\chi_q$.  
%
\subsubsection{$0\leq$ Im $\mE \leq$ Im  $\mEs$}
\begin{figure}
        \centerline{\includegraphics[width=8cm, height=6
cm]{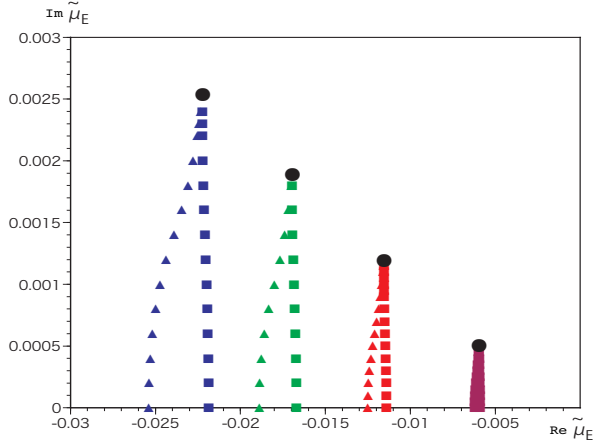}}
\caption{The $\tE$ dependence of the location  of the minimum  of Re $\Omega"$ (square symbol)  and  that  of $({\rm Re}~\chi_q)^{-1}$ (triangle)  in the complex $\mE$ plane.  Only the locations in  the region $0\leq$ Im~$\mE \leq$ Im~$\mEs$  for each $\tE$ are plotted. Filled black symbols indicate  $\mEs$ for four different temperatures $\tE=0.05$ (violet), 0.1 (red), 0.15 (green), 0.2 (blue). }
\label{maxchiq-minder2Om-fixim-tdep}
\end{figure}

For $\tE=0.1$, the left panel in Fig.~\ref{der2Omega-t01-fiximagall} indicates Re $\Omega"$ as a function of Re $\mE$ for fixed values of Im $\mE$ ($0\leq$ Im~$\mE \leq$ Im~$\mEs$).  Re~$\Omega"$ develops a minimum, which, as    ${\rm Im}~\mE \to {\rm Im}~\mEs$,   approaches zero at ${\rm Re}~\mE ={\rm Re}~\mEs$, where $\mEs=-0.011538+i \ 0.0011946$.   
Figure \ref{maxchiq-minder2Om-fixim-tdep}    indicates the location of the minimum  of Re~$\Omega"$ in the complex $\mE$ plane for various values of $\tE$ ($0.05, 0.1, 0.15, 0.2$). 
The minimum shows a    slight Im~$\mE$ dependence (the deviation between Re~$\mEs$  and Re~$\mE$ of  the  minimum point on the real axis is  at most  around 2 \% in the case under consideration).  
 As    ${\rm Im}~\mE \to {\rm Im}~\mEs$,   it approaches the singular point  $\mEs$, and  winds   around  a bit in the vicinity of $\mEs$, which reflects the behaviors of the   minima of Re~$\Omega$ around $\mEs$ (see Fig.~\ref{fig:extrema-traj-ttil0.1-rho}).   In contrast, the location of the minimum of the inverse  real  part of the  quark number susceptibility $({\rm Re}~\chi_q)^{-1}$ depend more on $\mE$ than that of  Re $\Omega"$  as shown in Fig.~\ref{maxchiq-minder2Om-fixim-tdep} (the deviation between Re~$\mEs$  and Re~$\mE$ of  the  minimum point on the real axis is  around 10 \% for $\tE=0.2$).  
 It is stressed   that  the crossover phenomena on the real axis    are originated from the same complex singularity, and that  $\chi_\sigma$ reflects the  singularity in the complex plane more directly  than  $\chi_q$ does.
  \par

\subsubsection{Im~$\mEs$ $\leq$ Im~$\mE$ }
\begin{figure}
        \centerline{\includegraphics[width=12cm, height=6
cm]{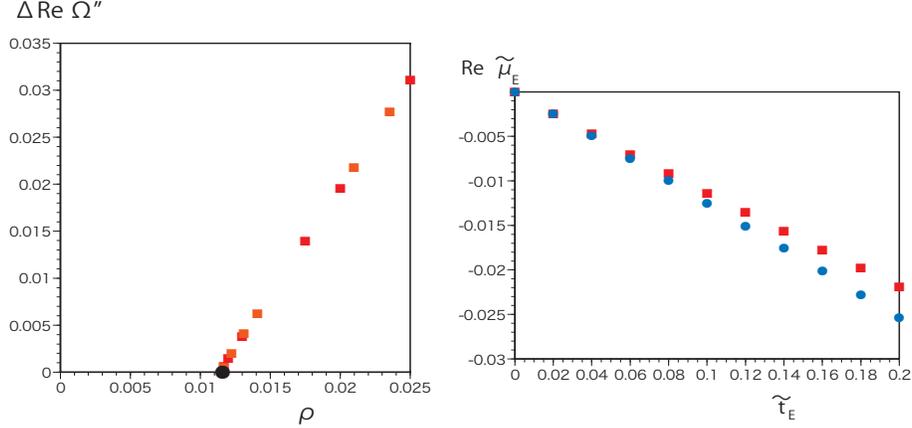}}
\caption{Left: The gap of  $ {\rm Re}~\Omega"$, $\Delta {\rm Re}~\Omega"$,   on the Stokes line,    increases  linearly  as a function of  distance $\rho$, where $\rho=\left| \mE \right|$ at a point on the Stokes line and $\rho^{(s)}=\left| \mEs \right|$ (filled black symbol).  $\tE=0.1.$  Right: The $\tE$ dependence of the location of the minimum of Re~$\Omega"$ (square)   and that   of $({\rm Re}~\chi_q)^{-1}$  (circle) on the real axis (Im~$\mE=0$). A linear fit works well  for both the quantities.}
\label{der2Omg-gap-im0-tdep}
\end{figure}

In the region Im~$\mEs$ $\leq$ Im~$\mE$, varying Re~$\mE$ comes across the Stokes line in the vicinity of $\mEs$. 
The right panel in Fig.~\ref{der2Omega-t01-fiximagall} indicates the behaviors of Re $\Omega"$ as a function of Re~$\mE$ for fixed values of Im~$\mE$.  Figure \ref{der2Omega-t01-fiximagall}  includes also those for  $0\leq$ Im~$\mE \leq$ Im~$\mEs$ as a comparison.   For  Im~$\mEs$ $\leq$ Im~$\mE$,  Re $\Omega"$ changes discontinuously at a value of Re $\mE$ corresponding to the Stokes line as shown in Fig.~\ref{der2Omega-t01-fiximagall}   (broken lines).   The gap of the  discontinuity  $\Delta {\rm Re}~\Omega"\equiv  {\rm Re}~\Omega"(\sigma_+)- {\rm Re}~\Omega"(\sigma_-)$ on the Stokes line  varies depending on how far the point is   from the singular point.  The left panel in Fig.~\ref{der2Omg-gap-im0-tdep} indicates that $\Delta {\rm Re}~\Omega"$ increases  linearly  as a function of  $\rho=\left| \mE \right|$ for $\tE=0.1$,  where $\mE$ is on the Stokes line.  At $\rho=\rho^{(s)}\equiv \left| \mEs \right|$,  $\Delta {\rm Re}~\Omega"$ is vanishing. 
\subsubsection{On the real axis}
 Let us turn to   behaviors  on the real $\mE$ axis. The right panel in Fig.~\ref{der2Omg-gap-im0-tdep} indicates the $\tE$ dependence of the location of the minimum of Re~$\Omega"$ (square) and that  of $({\rm Re}~\chi_q)^{-1}$ (circle) on the real $\mE$ axis.  It is seen that both the locations depend linearly on $\tE$  coming from the  linear    dependence of  Re~$\mEs$.   It is noted that  the difference between  Re~$\mEs$ and  the locations of the  minimum of Re~$\Omega"$ is invisibly small.\par
 The susceptibility in the complex plane  also behaves  with the critical  exponent characterized by the edge singularity.  On the real $\mE$ axis, it 
is given by $1/\delta-1=-2/3$, while 
in the complex plane,  it changes to  $-1/2$ (see Appendix \ref{ap-sus}). 
\section {Conclusion}
\label{conclusion}
We have discussed  the thermo-dynamic singularities in the complex chemical plane in  QCD  at  finite temperature and finite densities. For  this  purpose, we have adopted an effective theory incorporating fluctuations around  the CP.       Singularities  in the complex chemical potential plane are identified as    unstable points  of the  extrema  of the complex effective  potential. At  CP temperature, the singularity is located on the real $\mu$ axis, and   above CP temperature  it  moves away from the real axis leaving its  reminiscence   as a   crossover.  The location of the chiral susceptibility peak   agrees with   Re $\tilde \mu_E^{(s)}$ in the vicinity of the singularity.   \par
Simplicity  of the model allows us to explicitly deal with the complex potential as a function of the complex order parameter and complex values of $\mu$. 
  We have had  a close look at the behavior of  the extrema of ${\rm Re}\ \Omega(\sigma)$ in the complex order parameter plane.   It is seen that  two relevant extrema make a rearrangement at the singular point under the variation of $\mu$ around the singularity in the complex plane.     
It is also clearly seen that  the Stokes line is located in different ways depending on above, on and below CP temperature, which provide information as to where the Lee-Yang zeros are  located  for finite volume.  Along  the Stokes line, Im~$\Omega$ shows a gap, and the gap increases with the exponent characterized by  the Lee-Yang edge singularity.  \par
We have considered   the chiral  and quark number susceptibilities  in the complex plane.  
 As a  reminiscence of the singularity, the locations of the peaks of both the susceptibilities on the real 
 axis    exhibit  a linear dependence on $\tilde t_E$,   reflecting the $\tilde t_E$  dependence  of    Re~$\mEs$   in the complex plane.  The susceptibility in the complex plane  also behaves  with the critical  exponent characterized by the edge singularity.  
\par
Some remarks are in order.  (i)    The  framework discussed here  is not sufficient to make a realistic prediction by fixing some physical scale.  For this, a  study on more realistic model  beyond the  mean field framework  would be desirable.   (ii)  This model focuses on the vicinity of the TCP and CP,  and  physics  concerning the imaginary chemical potential  is beyond the scope of the present paper.    (iii) For  the sake of  concrete calculations  we have chosen some appropriate values for  the parameters (such as $C_a$ etc.)  appearing in the potential  in Eq. (\ref{eq:coeff}).      The universal behaviors discussed  in the complex plane are irrelevant to such a choice.   
(iv) It may be   worth while quantifying  what is described here.  
  In \cite{EY}, the QCD singularities $N_f=2$ QCD with staggered quarks have been investigated   by having a look at  the effective potential with respect to the plaquette variable.   In order to having a more contact with the present paper,  some refinement of the computation would be necessary. 
\begin{acknowledgments}
The authors are grateful  to   H. Kouno for useful discussion concerning the susceptibilities. They also thanks   H. Aoki,   M. Imachi and  M. Tachibana for useful discussion. 
S. E.  is in part supported by Grants-in-Aid of the Japanese Ministry of Education, Culture, Sports, Science and Technology (No. 23540259).  
\end{acknowledgments}


\appendix
\section{Stokes line}
\label{ap-Stokes}
In this appendix, we analytically consider the angle $\varphi$ of the  Stokes line  in the vicinity of the TCP, the critical point ($m=0$),  the first order phase transition point for $\tilde t_3<0$ ($m=0$)  and the CP ($\tilde t_E<0$), respectively.   
\subsection{Stokes line for the TCP}
  Firstly, we consider the case  $T=T_3$ ($\tilde t_3=0$).  
In   Eq.~(\ref{eq:TCP})  with $m=0$ and the coefficients in Eq.~(\ref{eq:linear}), 
the symmetric phase for $\tilde {\mu}_3 >0$ and the broken one for $\tilde {\mu}_3 <0$  compete in the vicinity of the TCP.  
In  the broken phase,   $\Omega$ develops two degenerate  minima for $|\tilde {\mu}_3| \ll 1$ such as 
\begin{eqnarray}
\bar \sigma\approx \pm e^{-i\pi/4}\sqrt{\frac{D_a}{c}} \tilde\mu_3^{1/4}=\pm (-\tilde\mu_3)^{1/4}, 
\end{eqnarray}
  while  $\bar \sigma=0$ in  the symmetric phase.
Therefore 
\begin{eqnarray}
\Omega(\tilde {\mu}_3)\ 
\left\{
\begin{array}{ccc}
\approx & -i\frac{D_a^{3/2}}{3\sqrt{c}}\tilde\mu_3^{3/2}= \frac{1}{3}(-\tilde\mu_3)^{3/2}  & (\tilde {\mu}_3<0)\\
=&  0 &  (\tilde {\mu}_3>0), 
\end{array}
\right.
\end{eqnarray}
where the second equality  for $\tilde {\mu}_3<0$ holds with   our  choice  of the parameters in Eq.(\ref{eq:parameters}).  Thus the condition  Eq.(\ref{eq2}) gives a tilted Stokes line with  angle $\varphi=\pi/3$. 
\subsection{Stokes line for the critical point  ($m=0$) }
\label{ap1}
In the massless case, the critical line runs for  $\tilde{t}_3 >0$, and the condition for this is $a=0$ in Eq.~(\ref{eq:TCP}), i.e., 
\begin{eqnarray}
\tilde \mu_3=-\frac{C_a}{D_a}\tilde t_3\equiv \tilde \mu_3^c. 
\end{eqnarray}
In the vicinity of $\mu_3^c$,  the global minimum and $\Omega$ behave as 
\begin{eqnarray}
\bar \sigma \ 
\left\{
\begin{array}{ccc}
\propto &  \frac{1}{\sqrt{\tilde t_3}}(\tilde\mu_3^c-\tilde\mu_3)^{1/2}  & (\tilde {\mu}_3<\tilde {\mu}_3^c)\\
=&  0 &  (\tilde {\mu}_3>\tilde {\mu}_3^c), 
\end{array}
\right. 
\end{eqnarray}
and
\begin{eqnarray}
\Omega(\tilde {\mu}_3)\  
\left\{
\begin{array}{ccc}
\propto &  -\frac{1}{\tilde t_3}(\tilde\mu_3^c-\tilde\mu_3)^{2}  & (\tilde {\mu}_3<\tilde {\mu}_3^c)\\
=&  0 &  (\tilde {\mu}_3>\tilde {\mu}_3^c),
\end{array}
\right.
\end{eqnarray}
respectively. 
The exponent of  $\Omega$  differs from that in the previous case ($\tilde t_3=0$), which  causes the change of  $\varphi$. 
The Stokes line    is thus  tilted  with  angle $\varphi=\pi/4$ in the vicinity of each critical point on the critical line. 
\subsection{Stokes lines for $\tilde{t}_3 <0$ ($m=0$)}
\label{ap2}
In the case of $\tilde{t}_3 <0$,  a first order phase transition occurs at $\tilde{\mu}_3=\mu_3^c$, where
\begin{eqnarray}
\mu_3^c=\frac{1}{3D_b^2}\left(8cD_a-3C_bD_b\tilde{t}_3-4\sqrt{4c^2D_a^2-3cD_b\left(C_bD_a-C_aD_b\right)\tilde{t}_3}\right). 
\end{eqnarray}
 In the vicinity of the $\mu_3^c$,  the global minimum and $\Omega$ behave as 
 \begin{eqnarray}
\bar \sigma \ =
\left\{
\begin{array}{ll}
  \bar \sigma_0+\bar \sigma_1(\tilde\mu_3-\tilde\mu_3^c)  & (\tilde {\mu}_3<\tilde {\mu}_3^c)\\
  0 &  (\tilde {\mu}_3>\tilde {\mu}_3^c), 
\end{array}
\right. 
\end{eqnarray}
and 
\begin{eqnarray}
\Omega(\tilde {\mu}_3)\  =
\left\{
\begin{array}{ll}
 d_1  (\tilde {\mu}_3-\tilde {\mu}_3^c)& (\tilde {\mu}_3<\tilde {\mu}_3^c)\\
  0 &  (\tilde {\mu}_3>\tilde {\mu}_3^c),
\end{array}
\right.
\end{eqnarray}
respectively,  
where $ \bar \sigma_0$, $\bar \sigma_1$ and $d_1$ are coefficients, which  depend  intricately  on $C_a$ etc.  in Eq.(\ref{eq:linear}).  The linear dependence  of $\Omega$ in the broken phase yields $\varphi=\pi/2$.  
It  suggests  in finite volume that the Lee-Yang zeros at  low temperatures and high densities are located parallel to the imaginary  $\mu$ axis. This  is in agreement with the recent Monte Carlo result obtained by utilizing the reduction formula of the reduced Dirac matrix~\cite{NMNNS}.
\subsection{Stokes line for the CP ($\tilde{t}_E =0$)}
\label{ap3}
At $\tilde{t}_E =0$ and  in the vicinity of the CP ($m\neq 0$),  the global minimum and $\Omega$ behave as follows.

\begin{eqnarray}
\bar \sigma \approx
\left\{
\begin{array}{ll}
  \sigma_0- \bar\sigma_1\ (-\tilde\mu_E)^{1/3}  & (\tilde {\mu}_E<0)\\
 \sigma_0- \bar\sigma_1\ (\tilde\mu_E)^{1/3} & (\tilde {\mu}_E>0),  
\end{array}
\right. 
\end{eqnarray}

\begin{eqnarray}
\Omega(\tilde {\mu}_E)\approx 
\left\{
\begin{array}{ll}
d_0\ (-\tilde\mu_E)^{4/3}  & (\tilde {\mu}_E<0)\\
d_0\ (\tilde\mu_E)^{4/3}  & (\tilde {\mu}_E>0).  
\end{array}
\right.
\end{eqnarray}
Here $\sigma_0$ is given in Eq.~(\ref{sig0}), and  $\bar \sigma_1$  and $d_0$ are coefficients, which  depend on $C_a$ etc.  in Eq.(\ref{eq:linear}).   In this case,  $\varphi=\pi/2$. 
\section{Critical exponent of the chiral susceptibility }
\label{ap-sus}
\begin{enumerate}
\item on the real $\tilde{\mu}_E$ axis ($\tE=0$)\\
Here, the behavior of the chiral susceptibility around the CP is discussed. 
Since the  singularity 
of the CP at $\tilde{t}_E=0$ is located at  the origin in the  complex $\tilde{\mu}_E$ plane, 
behaviors of the chiral susceptibility around the CP are given by fluctuations of $\sigma$ at the global 
minimum of $\Omega$ in the vicinity of $\tilde{\mu}_E^{(s)}=0$. 
\par
True vacuum $\bar{\sigma}$ which gives the global miminum of $\Omega(T,\mu,\sigma)$ is 
obtained by solving $\displaystyle{\frac{\partial\Omega}{\partial\sigma}=0}$, namely,
\begin{eqnarray}
	A_1+2A_2\hat{\sigma}+3A_3\hat{\sigma}^2+4A_4\hat{\sigma}^3=0.
\end{eqnarray}
 This equation is cubic and it can be analytically solved. Ignoring higher-order terms 
 in $\tilde{\mu}_E$ around the CP, one obtains 
\begin{eqnarray}
	\bar{\sigma}\simeq   \sigma_0+ \left(\frac{D_a\sigma_0+D_b\sigma_0^3}{2b(T_E,\mu_E)}\right)^{1/3}
	|\tilde{\mu}_E|^{1/3}.
\end{eqnarray}
\par
The curvature of $\Omega$ for true vacuum is calculated as follows:
\begin{eqnarray}
	\frac{\partial^2\Omega(\bar{\sigma}(\tilde{\mu}_E),\mu)}{\partial\sigma^2}
	\simeq 
	-6b(T_E,\mu_E)\left(\frac{D_a\sigma_0+D_b\sigma_0^3}{2b(T_E,\mu_E)}\right)^{2/3}
	|\tilde{\mu}_E|^{2/3},
\end{eqnarray}
where  the contributions of the first term  ($\mathcal O (|\tilde{\mu}_E|^{1})$ ) and second  one  ($\mathcal O (|\tilde{\mu}_E|^{4/3})$)    are neglected due to $|\tilde{\mu}_E|\ll 1$.
\par
The chiral susceptibility, thus, behaves as
\begin{eqnarray}
	\chi_{\sigma}\sim |\tilde{\mu}_E|^{-2/3}
\end{eqnarray}
in the vicinity of the origin in the complex $\tilde{\mu}_E$  plane.
\item  in the complex  $\tilde{\mu}_E$ plane ($\tE>0$)  \\ 
For $\tE>0$ with fixed Im~$\mE=$ Im~$\mEs$, $\bar\sigma$ behaves like
\begin{eqnarray}
\delta \bar\sigma\equiv \bar\sigma-\sigma^{(s)}\sim \left(\delta \mE\right)^{1/2},
\end{eqnarray}
 in the vicinity of the singular point ($\delta \mE\equiv \mE-\mEs$).  
Similarly to the $\tE=0$ case,  $ \delta \Omega"\equiv \Omega"(\bar\sigma)- \Omega"(\sigma^{(s)}) \ ( \Omega"(\sigma^{(s)})=0)$  behaves like  
\begin{eqnarray}
  \delta \Omega" &=&
  2\delta A_2+6\delta \left(A_3\hat\sigma\right)+12 \delta \left(A_4\hat\sigma^2\right)\nonumber \\
 &=& 
 D_a \left(\delta \mE\right)^1+6\sigma_0D_b  \sigma^{(s)}\left(\delta \mE\right)^1+6\sigma_0D_b \mEs\left(\delta \mE\right)^{1/2}
 +12(-b/2)2\sigma^{(s)}\left(\delta \mE\right)^{1/2}  \nonumber \\
 &\sim & \left(\delta \mE\right)^{1/2}, 
\end{eqnarray}
leading to 
\begin{eqnarray}
{\rm Re}~\chi_{\sigma} \sim \left(\delta \mE\right)^{-1/2}. 
\end{eqnarray}
\end{enumerate}
\newpage 

\end{document}